\documentclass[aps, showpacs, showkeys,nofootinbib,floatfix]{revtex4}

\usepackage{amssymb}
\usepackage{amsmath}
\usepackage{graphicx}

\newcommand{\beq}[1]{\begin{equation}\label{#1}}
 \newcommand{\eeq}{\end{equation}}
 \newcommand{\bea}{\begin{eqnarray}}
 \newcommand{\eea}{\end{eqnarray}}

\newcommand{\aap}{Astron. and Astrophys. }
\newcommand{\apjs}{Astrophys. J. Suppl. }
\newcommand{\mnras}{Mon. Not. Roy. Astron. Soc. }
\newcommand{\jcap}{JCAP }
\newcommand{\aj}{Astron. J. }

\def\la{\mathrel{\hbox{\rlap{\hbox{\lower4pt\hbox{$\sim$}}}\hbox{$<$}}}}

\begin{document}


\title
{Testing the interaction model with cosmological data and gamma-ray bursts}

\author{Yu Pan$^{1,2}$, Shuo Cao$^{1}$, Yungui Gong$^{3}$, Kai Liao$^{1}$, Zong-Hong Zhu$^{1}$}
\email{zhuzh@bnu.edu.cn}
\affiliation{ $^1$ Department of Astronomy, Beijing Normal
University, Beijing 100875, China \\
$^2$ College of Mathematics and Physics, Chongqing Universe of Posts
and Telecommunications, Chongqing 400065, China \\
$^3$ MOE Key Laboratory of Fundamental Quantities Measurement, School of Physics, Huazhong University of Science and Technology, Hubei 430074, China}







\begin{abstract}
We use the new gamma-ray bursts (GRBs) data, combined with the baryon acoustic oscillation(BAO) observation from the 
spectroscopic Sloan Digital Sky Survey (SDSS) data release, the newly obtained $A$ parameter at $z=0.6$ from the WiggleZ Dark Energy Survey, 
the cosmic microwave background (CMB) observations from the 7-Year Wilkinson Microwave Anisotropy Probe (WMAP7) results, and the type Ia supernovae
(SNeIa) from Union2 set, to constrain a
phenomenological model describing possible interactions between dark energy and dark matter, which was proposed to alleviate the coincidence problem of the standard $\Lambda$CDM model.
By using the Markov Chain Monte Carlo (MCMC) method, we obtain the marginalized $1\sigma$ constraints $\Omega_{m}=0.2886\pm{0.0135}$, $r_m=-0.0047\pm{0.0046}$, and $w_X=-1.0658\pm{0.0564}$.
We also consider other combinations of these data for comparison. These results show that: (1) the energy of dark matter is slightly transferring to that of dark energy;
(2) even though the GRBs+BAO+CMB data present less stringent constraints than SNe+BAO+CMB data do, the GRBs can help eliminate the degeneracies among parameters.

\end{abstract}


\maketitle
\section{Introduction}\label{sec1}
A lot of astrophysical and cosmological observations have indicated that the universe is undergoing an accelerating expansion and great efforts have been made to understand the driving force behind the cosmic acceleration \citep{Riess98,Perlmutter99,Astier06,Hicken09,Amanullah10,Spergel03,Spergel07,Komatsu09,Komatsu11,Tegmark04,Eisenstein05,Cao12a,Cao12b,Cao11a,Gong12}. This gives birth to the construction of a strange dark energy with negative pressure, which may contribute to interpret the present accelerated expansion. The most simple candidate for these uniformly distributed dark energy is considered to be in the form of vacuum energy density or cosmological constant ($\Lambda$). However, despite its simplicity, the simple cosmological constant is always entangled with the coincidence problem: Why the present matter density $\rho_m$, which
decreases with the expansion of our universe with $a^{-3}$, is comparable with the dark energy density $\rho_\Lambda$, which does not change with the cosmic expansion of our universe? In order to relieve the coincidence problem, other dynamic dark energy models were proposed in the past decades, including quintessence \citep{Ratra88, Caldwell98}, phantom \citep{Caldwell02,Caldwell03}, k-essence \citep{Armendariz-Picon01, Chiba02}, as well as quintom model \citep{Feng05, Feng06, Guo05a}. However, the nature of dark energy is still unknown. The other presumption is naturally considered that energy is exchanged between dark energy and dark matter through interaction.

We assume that dark energy and dark matter exchange energy through interaction term $Q$, namely
\bea
 &&\dot{\rho}_X+3H\rho_X(1+w_X)=-Q,\label{eq.eosX}\\
 &&\dot{\rho}_m+3H\rho_m=Q\label{eq.eosM},
 \eea
and the total energy conservation equation expresses as $\dot{\rho}_{tot}+3H\left(\rho_{tot}+p_{tot}\right)=0$,
where $\rho_{tot}=\rho_{X}+\rho_{m}$. Because the format of interaction term  still can not be determined from fundamental physics, many literature have extensively considered various forms of the interaction term $Q$ \citep{Wei06,Wei05a,Alimohammadi06,Coley03,Copeland98,Amendola99, Guo05b,Ellis89,Wei05b,Wei07a,Zimdahl01,Cai05,Szydlowski06}

In this work, we consider the simplest case for convenience, i.e. \citep{Wei07b}
\beq{eq.modelQ}
Q=3r_mH\rho_m,
\eeq
where $r_m$ is a dimensionless constant: $r_m=0$ indicates that there is no interaction between dark energy and dark matter; the energy is transferred from dark matter to dark energy when $r_m<0$, and from dark energy to dark matter when $r_m > 0$.  We assume that the equation of state (EoS) of dark energy, $w_X\equiv{p_X}/\rho_{X}$, is a constant in a spatially flat FRW metric. In this case, Eq.~(\ref{eq.eosX}) reads
\beq{eq.tmpXM}
 \dot{\rho}_X+3H\rho_X(1+w_X)=-3r_mH\rho_m.\\
\eeq
Combining Eq.~(\ref{eq.eosM}) and Eq.~(\ref{eq.tmpXM}), we can get
\beq{eq.rhoM}
\rho_{m}=\rho_{m0}(1+z)^{3(1-r_m)}=\rho_{m0}a^{-3(1-r_m)}
\eeq
By using Eq.~(\ref{eq.rhoM}), we obtain
\beq{eq.rhoX}
\rho_X=Aa^{-3(1+w_X)}-\frac{r_m\rho_{m0}}{(r_m+w_X)}a^{-3(1-r_m)},
\eeq
where $A$ is an integral constant. Inserting it into the Friedmann equation
\beq{eq.tmpfried}
H^2=\kappa^2(\rho_m+\rho_X)/3,
\eeq
where $\kappa^2=8\pi{G}$ with $G$ the gravitational constant, and this integral constant can determined with $H(z=0)=H_0$. The corresponding Hubble parameter is
\beq{eq.fried}
E^2(z)=(H/H_0)^2=\frac{w_X\Omega_{m}}{(r_m+w_X)}(1+z)^{3(1-r_m)}+\\
(\frac{1-w_X\Omega_{m}}{(r_m+w_X)})(1+z)^{3(1+w_X)}.
\eeq

Recently the simple phenomenological interacting scenario has been constrained from several cosmological observations \citep{Wei07a,Cao11b,Chen10}. Also, the Gamma-ray bursts(GRBs) have been proposed as distance indicators and regarded as a complementary cosmological probe of the universe at high redshifts \citep{Schaefer03,Dai04,Ghirlanda04,Firmani05,Firmani06,Liang05,Ghirlanda06,Schaefer07,Wang07,Wright07,Amati08,Basilakos08,Mosquera08a,Mosquera08b,Daly08}. In contrast to supernovae, the high energy photons in the gamma-ray band are nearly unaffected by dust extinction. Therefore, those observed high-redshift Gamma-Ray Bursts (GRBs) at $0.1<z<8.1$ may constitutes are a complementary probe to fill the "desert" between the redshifts of SNIa and CMB. Nevertheless, due to the lack of a low-redshift GRBs at $z<0.1$, there is a circularity problem in the direct use of GRBs \citep{Ghirlanda06}. Some statistical methods have been proposed to alleviate this problem, such as the scatter method \citep{Ghirlanda04}, the luminosity distance method \citep{Ghirlanda04} and the Bayesian method \citep{Firmani05}. Other methods trying to avoid the circularity problem have been proposed in Ref.~\citep{Li08,Liang06}. Using the cosmology-independent calibration method proposed in Ref.~\citep{Liang08}, \citep{Wei10} have obtained 59 calibrated high-redshift GRBs called "Hymnium" GRBs sample out of 109 long GRBs with the well-known Amati relation. Therefore, it may be rewarding to test the coupling between dark sectors with this newly obtained 59 GRBs deprived of the circularity problem.

With this aim, in this paper, we adopt the Markov Chain Monte Carlo(MCMC) technique to constrain one interaction model from the latest observational data. To reduce the uncertainty and put tighter constraint on the value of the coupling, we combine the GRBs data with the joint observations such as the 557 Union2 SNeIa dataset \citep{Amanullah10}, the CMB observation from the Wilkinson Microwave Anisotropy Probe (WMAP7) \citep{Komatsu11} results, and the model independent new $A$ parameter from the baryon acoustic oscillation (BAO) measurements \citep{Blake11}, the two BAO distance ratios at $z=0.2$ and $z=0.35$ from the spectroscopic Sloan Digital Sky Survey (SDSS) data release7 (DR7) galaxy sample \citep{Percival10}. This paper is organized as follows: we introduce the observational data in Section \ref{sec.data}. The numerical analysis results are discussed in Section \ref{sec.result}, and the main conclusions are summarized in Section \ref{sec.conclusion}.

\section{Observational data}\label{sec.data}

The 59 calibrated Hymnium GRBs and the 557 Union2 SNeIa data sets are given in term of the distance modulus $\mu(z)$. Theoretically, the distance modulus can be calculated as
\beq{eq.mu}
\mu=5\log\frac{d_L}{Mpc}+25=5\log_{10}H_0d_L-\mu_0,
\eeq
where $\mu_0=5\log_{10}[H_0/(100km/s/Mpc)]+42\cdot38$, and the luminosity distance $d_L$ can be calculated using $d_L=\frac{(1+z)}{H_0}\int_0^z\frac{dz'}{E(z')}$. The $\chi^2$ value of the observed distance moduli can be calculated by
\beq{eq.tmpchiSN}
\chi^2_\mu=\sum_{i=1}^{N}\frac{[\mu(z_i)-\mu_{obs}(z_i)]^2}{\sigma_{\mu{i}}^2},
\eeq
where $\mu_{obs}(z_i)$ is the observed distance modulus for the SNe Ia or
GRBs at redshift $z_i$ with error $\sigma_{\mu{i}}$; $\mu(z_i)$ is the theoretical value of the distance modulus calculated from Eq. (\ref{eq.mu}). The nuisance parameter $h$ is marginalized with a flat prior, after which we get \citep{Gong08}
\beq{eq.chiSN}
\chi^2_\mu=\sum_{i=1}^{N}\frac{\alpha_i^2}{\sigma^2_\mu{i}^2}-\frac{(\sum_{i=1}^N\alpha_i/\sigma_{\mu{i}}^2-\ln 10/5)^2}{\sum_{i=1}^N 1/\sigma_{\mu{i}}^2}\nonumber\\
-2\ln\left(\frac{\ln 10}{5}\sqrt{\frac{2\pi}{\sum_{i=1}^N  1/\sigma_i^2}}\right),
\eeq
where $\alpha_i=\mu_{obs}(z_i)-25-5\log_{10}H_0d_L$. In this part the radiation component of the total density is neglected, because its contribution in low redshifts is negligible.

For the BAO observation, we use the new BAO $A$ parameter at $z=0.6$, with the measured value $A=0.452\pm{0.018} $ \citep{Blake11} from the WiggleZ Dark Energy Survey. It can be calculated as the following:
\begin{equation}
A=\sqrt{\Omega_{m}}\frac{H_0 D_V(z=0.6)}{z=0.6}\\
 =\frac{\sqrt{\Omega_{m}}}{0.6}\left[\frac{0.6}{E(0.6)}\frac{1}{|\Omega_{k}|}
{\rm{sinn}}^2\left(
\sqrt{|\Omega_{k}|}\int_0^{0.6}\frac{dz}{E(z)}\right)\right]^{1/3},
\end{equation}
 Now we can add our $\chi^2$ obtained before with
\begin{equation}
\chi^2_{BAOa}(\mathbf{p})=\left(\frac{A-0.452}{0.018}\right)^2.
\end{equation}
Notice that the BAO A parameter does not depend on the baryon density $\Omega_bh^2$ or the Hubble constant $h$. The radiation density which does depend on $h$ is negligible to the Hubble parameter $E(z)$.

In addition to the above $A$ parameter, we also consider the BAO distance ration $(d_z)$ at $z=0.2$ and $z=0.35$ from SDSS data release 7 (DR7) galaxy sample \citep{Percival10}. The BAO distance ration can be expressed as
\begin{equation}
\label{dz} d_{z}= \frac{r_{s}(z_{d})}{D_{V}(z)},
\end{equation}
where the effective distance is given by \citep{Eisenstein05}
\begin{equation}
\label{dvdef}
D_V(z)=\left[\frac{d_L^2(z)}{(1+z)^2}\frac{z}{H(z)}\right]^{1/3};
\end{equation}
and the drag redshift $z_d$ is fitted as \citep{Eisenstein98}
\begin{equation}
\label{zdfiteq} z_d=\frac{1291(\Omega_m
h^2)^{0.251}}{1+0.659(\Omega_m h^2)^{0.828}}[1+b_1(\Omega_b
h^2)^{b_2}],
\end{equation}
\begin{eqnarray}
\label{b1eq} b_1=0.313(\Omega_m h^2)^{-0.419}[1+0.607(\Omega_m
h^2)^{0.674}],\nonumber\\
\quad b_2=0.238(\Omega_m h^2)^{0.223},
\end{eqnarray}
The comoving sound horizon is
\begin{equation}
\label{rshordef} r_s(z)=\int_z^\infty \frac{c_s(z)dz}{E(z)},
\end{equation}
where the sound speed $c_s(z)=1/\sqrt{3[1+\bar{R_b}/(1+z)}]$, and $\bar{R_b}=3\Omega_b h^2/(4\times2.469\times10^{-5})$.
The $\chi^2$ value of BAO observation can be expressed as \citep{Percival10}
\begin{eqnarray}
\chi^2_{\mathrm{BAO}}=\Delta
\bf{P}_{\mathrm{BAO}}^\mathrm{T}{\bf
C_{\mathrm{BAO}}}^{-1}\Delta\bf{P}_{\mathrm{BAO}},
\end{eqnarray}
where $\bf{\Delta{P}_{BAO}}=\bf{P}_{th}-\bf{{P}_{obs}}$, $\bf{{P}_{obs}}$ is the observed distance ration. ${\bf{C_{\mathrm{BAO}}}}^{-1}$ is the corresponding inverse covariance matrix.

For the CMB observation, we use the derived dataset from the WMAP7 measurement, including the acoustic scale $(l_a)$ ,the shift parameter $R$, and the redshift of recombination $z_\ast$ \citep{Komatsu11}. The acoustic scale can be expressed as
\begin{equation}
l_a=\pi\frac{\Omega_\mathrm{k}^{-1/2}sinn[\Omega_\mathrm{k}^{1/2}\int_0^{z_{\ast}}\frac{dz}{E(z)}]/H_0}{r_s(z_{\ast})},
\end{equation}
where $r_s(z_{\star})
={H_0}^{-1}\int_{z_{\ast}}^{\infty}c_s(z)/E(z)dz$ is the comoving
sound horizon at photon-decoupling epoch. The shift parameters can be
expressed as
\begin{equation}
R(z_\ast)=\frac{\sqrt{\Omega_{m}}}{\sqrt{|\Omega_{k}|}}{\rm
sinn}\left(\sqrt{|\Omega_{k}|}\int_0^{z_\ast}\frac{dz}{E(z)}\right).
\end{equation}

The decoupling redshift $z_\ast$ is fitted by \citep{Hu96},
\begin{eqnarray}
\label{zasteq} z_\ast=1048[1+0.00124(\Omega_b
h^2)^{-0.738}][1+g_1(\Omega_m h^2)^{g_2}],
\end{eqnarray}
\begin{equation}
g_1=\frac{0.0783(\Omega_b h^2)^{-0.238}}{1+39.5(\Omega_b
h^2)^{0.763}},\quad g_2=\frac{0.560}{1+21.1(\Omega_b h^2)^{1.81}}.
\end{equation}

The $\chi^2$ for CMB can be expressed then as
\beq{eq.chi_cmb}
\chi^2_{\mathrm{CMB}}=\Delta
\textbf{P}_{\mathrm{CMB}}^\mathrm{T}{\bf
C_{\mathrm{CMB}}}^{-1}\Delta\textbf{P}_{\mathrm{CMB}},
\eeq
where $\Delta\bf{P_{\mathrm{CMB}}} =
\bf{P_{\mathrm{th}}}-\bf{{P}_{\mathrm{obs}}}$, $\bf{{P}_{\mathrm{obs}}}$ is the observed dataset, and the
${\bf
C_{\mathrm{CMB}}}^{-1}$ is corresponding inverse covariance matrix.

\section{Constraint on the phenomenological interacting scenario}\label{sec.result}
The model parameters are determined by applying the minimum likelihood method of $\chi^2$ fit. Basically, the model parameters are determined by minimizing
\bea
\chi^2=\chi^2_{GRBs}+\chi^2_{SNe}+\chi^2_{CMB}+\chi^2_{BAO}+\chi^2_{BAOa}.
\eea

\begin{figure}
\begin{center}
\includegraphics[width=0.6\hsize]{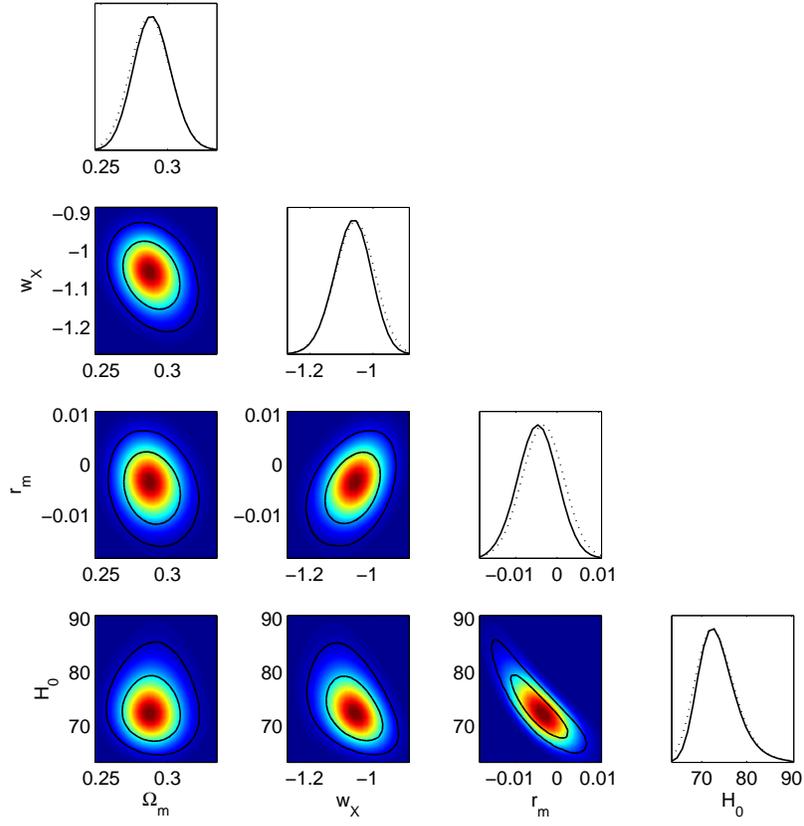}
\end{center}
\caption{The marginalized probability distributions, and the marginalized $1\sigma$ and $2\sigma$ confidence contours
of the parameters $w_X$, $r_m$, $\Omega_{m}$, and $H_0$ in the phenomenological interacting scenario, from the combinations of all observational data
SNe+BAO+CMB+GRBs. \label{fig.ct_all}}
\end{figure}

\begin{figure}
\begin{center}
\includegraphics[width=0.4\hsize]{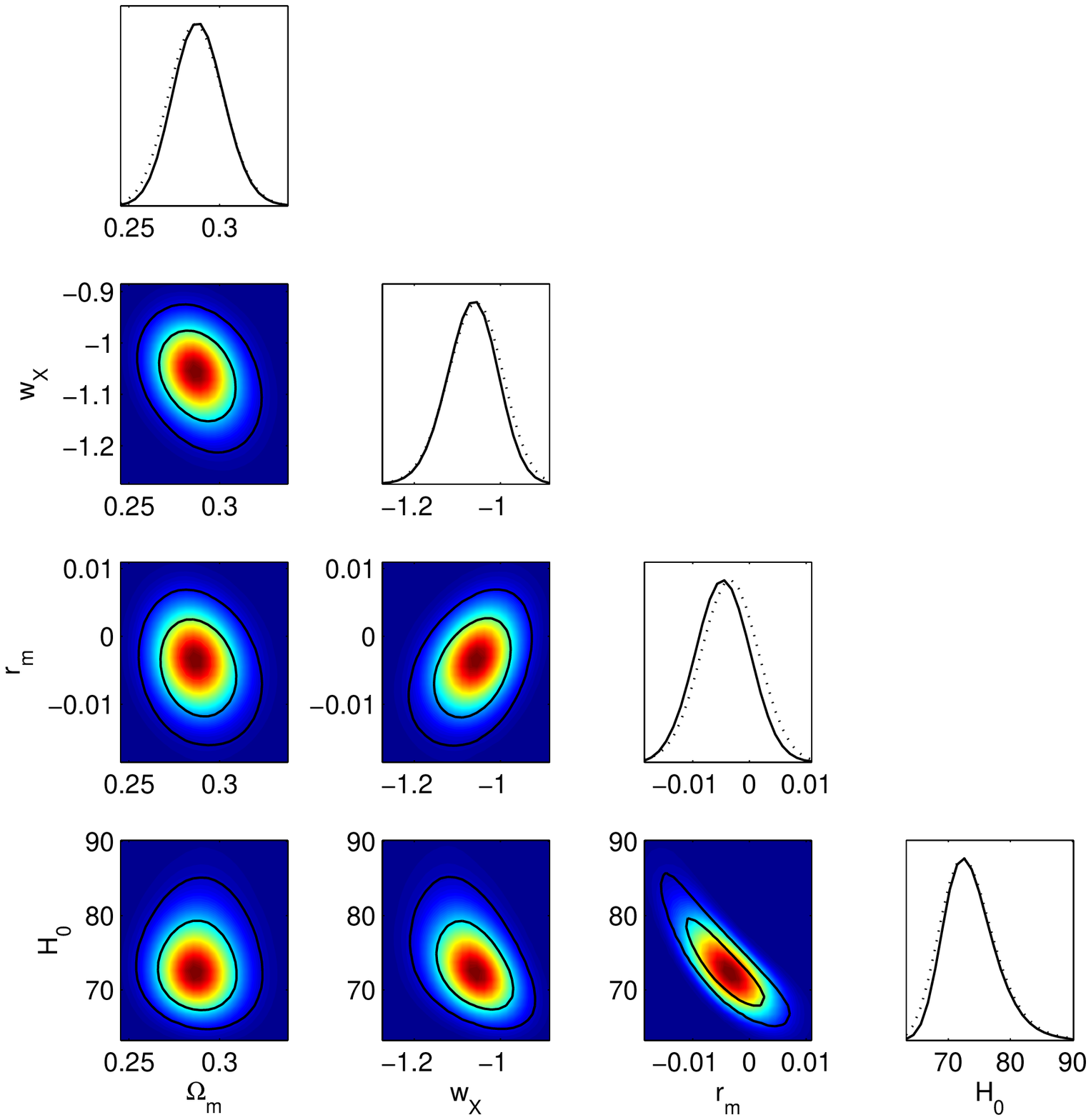} \includegraphics[width=0.4\hsize]{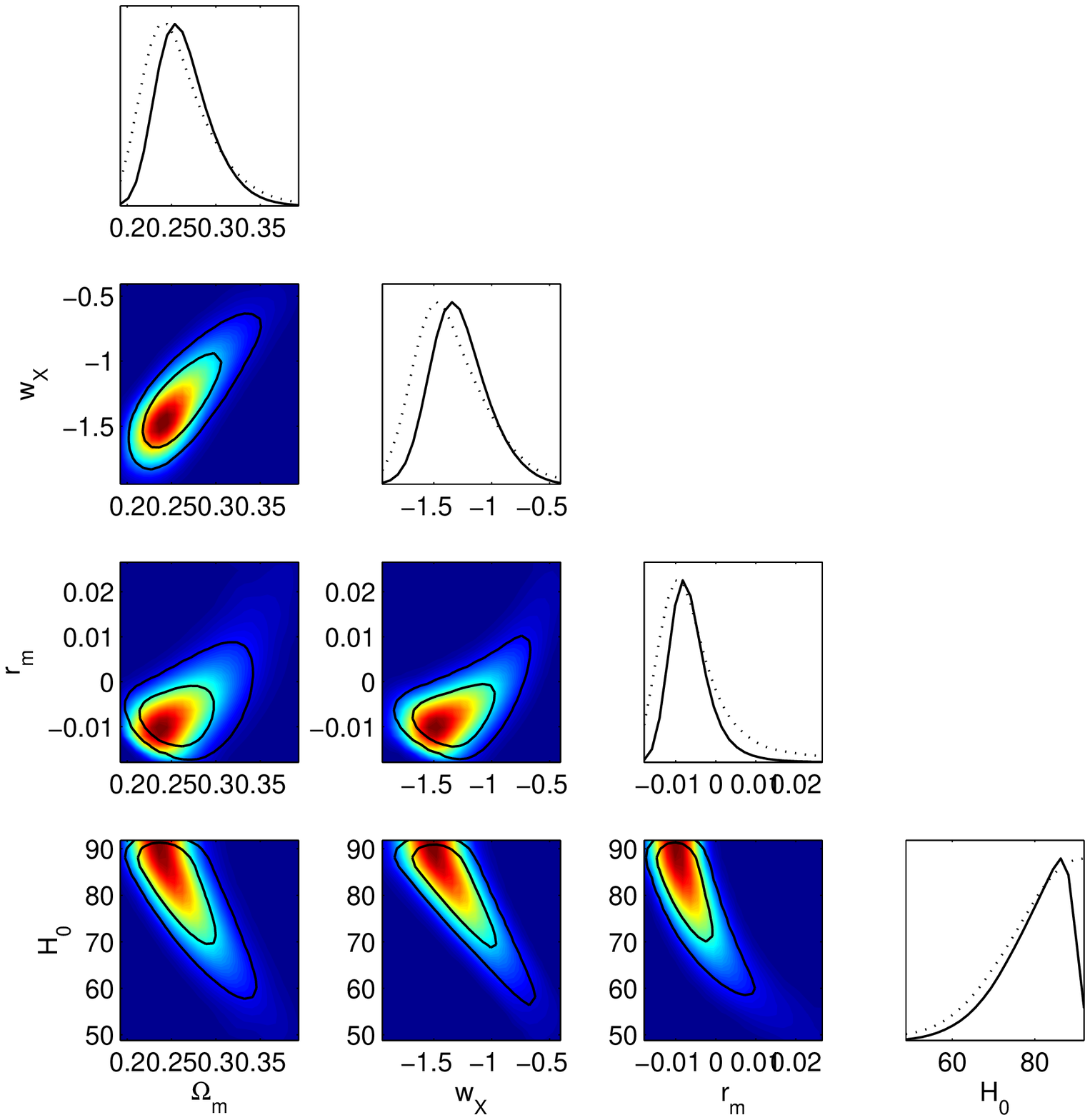}
\caption{The same as in Figure \ref{fig.ct_all}, but the constraints in the left panel are from the data combinations SNe+BAO+CMB (all-GRBs) and
the constraints in the right panel are from the data combinations GRBs+BAO+CMB (all-SNe). \label{fig.ct_snetc}}
\end{center}
\end{figure}

We apply the Monte Carlo Markov Chain (MCMC) method \citep{Lewis02} with 8 chains and obtain the marginalized $1\sigma$ constriants:
$\Omega_{m}= 0.2886\pm{0.0135}$,
$r_m=-0.0047\pm{0.0046}$, and
$w_X=-1.0658\pm{0.0564}$. We show the marginalized probability distribution of each parameter and the marginalized 2D confidence contours of parameters in Figure \ref{fig.ct_all}.
For comparison, fitting results from the combinations of SNe+BAO+CMB and GRBs+BAO+CMB are shown in Figure \ref{fig.ct_snetc}.
Because CMB and BAO data tightly constrain the matter density $\Omega_{m}$, and therefore help improve the constraints on dark energy property \cite{Gong12},
so these data are taken as priors in our treatment and are combined with other data to test the constraining power of GRBs and SNe data.
It is shown that the plots in Figure \ref{fig.ct_all} and the left panel of Figure \ref{fig.ct_snetc} are almost the same,
suggesting that the current GRBs data are consistent with other observations, although they contribute little to the existing constraints.
Comparing the constraints from SNe data with those from GRBs data in Figure \ref{fig.ct_snetc}, we see that the degeneracies between $\Omega_{m}$ and $w_X$, $\Omega_{m}$ and $r_m$ are
different although the current GRBs data give larger errors on $w_X$. Therefore, GRBs data have the potential to help constrain the model parameters $w_X$ and $r_m$ by combining with SNe data.
The best-fit values of the parameters along with their $1\sigma$ uncertainties from all three different combinations mentioned above, are explicitly presented in Table \ref{tab.interval}.
In general, for the three different joint data sets, at the $1\sigma$ confidence region, the energy is seen transferred from dark matter to dark energy, and
the concordance $\Lambda$CDM $(w_X=-1,r_m=0)$ model can not be excluded.

From Figure \ref{fig.ct_all}, we see that the coupling parameter $r_m$ is correlated with all other parameters $\Omega_{m}$, $w_X$ and $H_0$. However, $r_m=0$ is within
the $1\sigma$ confidence region. To see the point clearly, we fix $r_m$=0 and obtain the marginalized $1\sigma$ uncertainties of the other model parameters from the combination of all
observational data (SNe+CMB+BAO+GRBs): $\Omega_{m}=0.2863\pm{0.0130}$, $w_X=-1.0436\pm{0.0509}$, and ${H_0}/100=0.6994\pm{0.0118}$.
The constraints on the model without interaction ($r_m=0$) are consistent with those with interaction at $1\sigma$ level, and $w_X$ shifts toward $-1$ when $r_m=0$ so that it
is consistent with $\Lambda$CDM model at $1\sigma$ level.

\begin{table*}
 \begin{center}{\scriptsize
 \begin{tabular}{|c|c|c|c|c|c|} \hline\hline
 \cline{1-6}
  Data  \ \ & \ \ $w_X$   &$r_m$   &$\Omega_{m}$  &${H_0}/100$  &$\chi^2$/dof\\ \hline
 all \ \ & \ \ $-1.0658\pm{0.0564}(1\sigma)$ \ \ & \ \ $-0.0047\pm{0.0046}(1\sigma)$\ \ & \ \ $0.2886\pm{0.0135}(1\sigma)$\ \ & \ \  $0.7369\pm{0.0391}(1\sigma)$\ \ & \ \ $0.9140$\ \\ \hline
all-GRBs \ \  & \ \ $-1.0657\pm{0.0562}(1\sigma)$ \ \ & \ \ $-0.0047\pm{0.0046}(1\sigma)$ \ \ & \ \ $0.2886\pm{0.0135}(1\sigma)$ \ \  & \ \ $0.7367\pm{0.0389}(1\sigma)$ \ \ & \ \ $0.9686$ \ \ \\ \hline
all-SNe \ \ & \ \ $-1.2757\pm{0.2358}(1\sigma)$ \ \ & \ \ $-0.0062\pm{0.0045}(1\sigma)$ \ \   & \ \ $0.2649\pm{0.0288}(1\sigma)$ \ \ & \ \ $0.7984\pm{0.0803}(1\sigma)$ \ \ & \ \ $0.3580$ \ \ \\ \hline\hline
 \end{tabular}}
 \end{center}
 \caption{The marginalized $1\sigma$ errors of the parameters $w_X$, $r_m$, $\Omega_{m}$, and $H_0$ for the phenomenological scenario, as well as $\chi^2$/dof, obtained from
 the combinations of the data sets GRBs+SNe+BAO+CMB (all), SNe+BAO+CMB (all-GRBs), and GRBs+BAO+CMB (all-SNe). \label{tab.interval}}
 \end{table*}

\section{Conclusions}\label{sec.conclusion}

In this paper we constrain an interacting dark energy model using the new GRBs, the Union2 SNe, CMB, and BAO data sets.
By adopting the MCMC approach we obtained the marginalized $1\sigma$ errors of each parameter: $\Omega_{m}=0.2886\pm{0.0135}$,
$r_m=-0.0047\pm{0.0046}$, and
$w_X=-1.0658\pm{0.0564}$.

In order to test the constraining power of the observational GRBs data, we compared the results from the combinations
of GRBs+BAO+CMB with those from SNe+BAO+CMB. We find that the constraints from the current GRBs data are less stringent than those from SNe data.
The error bars of $w_X$, $\Omega_m$ and $H_0$ from GRBs+BAO+CMB are roughly twice of those from SNe+BAO+CMB, although the error bar
of $r_m$ is roughly the same as shown in Table \ref{tab.interval} and Figure \ref{fig.ct_snetc}.
However, the directions of degeneracies between $\Omega_{m}$ and $w_X$, $\Omega_{m}$ and $r_m$ are
different, so GRBs data have the potential to help tighten the constraint on the parameters $\Omega_m$, $w_X$ and $r_m$
if the measurement precision of the data is improved in the future.
By fitting the model to all the observational data combined, we find that the energy slightly transfers from dark matter to dark energy at ${1\sigma}$ region. We also note that the coupling parameter $r_m$ is correlated with all other model parameters $\Omega_m$, $w_X$ and $H_0$,
and $r_m=0$ is within the ${1\sigma}$ confidence region. The constraints on the model without interaction ($r_m=0$) are consistent with those with interaction at $1\sigma$ level, and the value of $w_X$ shifts up a little when $r_m$=0. In conclusion,
the concordance $\Lambda$CDM model still remains a good fit to the observational data.

\section*{Acknowledgments}
We are grateful to Hao Wang, Yun Chen, Xiao-long Gong, Ling-Zhi Wang, Xing-jiang Zhu, Yan Dai, Fang Huang, Jing Ming, Kai Liao, Yubo Ma, Huihua Zhao, Yi Zhang, and Na-Na Pan for helpful discussions. This work was supported by the National Natural Science
Foundation (NNSF) of China under the Distinguished Young Scholar
Grant 10825313, the NNSF of China under grant Nos. 10935013 and 11175270 ,the Ministry of Science and Technology national basic science Program (Project 973) under grant No.
2012CB821804, the Fundamental Research Funds for the Central Universities and Scientific Research Foundation of Beijing
Normal University, the Excellent Doctoral Dissertation of Beijing Normal University Engagement Fund,
CQ CSTC under grant No. 2009BA4050, CQ CMEC under grant Nos. KJTD201016 and KJ110523,
and the Nature Science Foundation of Chongqing University of Posts and Telecommunications under grant No.A2011-27.



\end{document}